\documentclass[twocolumn,showpacs]{revtex4}

\input{epsf}

 \def\CQG{{\it Class. Quantum
Gravity} }
    \def\IJMP{{\it Int. J. Mod. Phys.} }

  \def\PR{{\it
Phys. Rev.} } \def\PRL{{\it Phys. Rev. Lett.} }

\def\al{\alpha}   
\def\ep{\epsilon}   
\def\th{\theta}

 \def\Om{\Omega}

 \def\frac#1#2{{\textstyle{{#1}\over
{#2}}}} 
\def\lsim{\mathrel{\rlap{\lower4pt\hbox{\hskip1pt$\sim$}}
\raise1pt\hbox{$<$}}}
\def\gsim{\mathrel{\rlap{\lower4pt\hbox{\hskip1pt$\sim$}}
\raise1pt\hbox{$>$}}} \def\sqr#1#2{{\vcenter{\vbox{\hrule
height.#2pt \hbox{\vrule width.#2pt height#1pt \kern#1pt \vrule
width.#2pt} \hrule height.#2pt}}}}

\def\beq{\begin{equation}} \def\eeq{\end{equation}}
\def\beqa{\begin{eqnarray}} \def\eeqa{\end{eqnarray}}

\def\@fnsymbol#1{\ifcase#1\hbox{}\or *\or \dagger\or \ddagger\or \mathchar "278\or \mathchar "27B\or \|\or **\or \dagger\dagger \or \ddagger\ddagger \or \mathchar"27C \else\@ctrerr\fi\relax}

\long\def\symbolfootnote[#1]#2{\begingroup
\def\thefootnote{\fnsymbol{footnote}}\footnote[#1]{#2}\endgroup}

\newcommand{\mr}[1]{\mathrm{#1}}

\begin{document}

\title{Thermal Analysis of the Pioneer Anomaly: \\A method to estimate radiative momentum transfer}

\vskip 0.2cm

\author{O.~Bertolami$^{1,3}$\footnotetext{$^1$ Departamento de
F\'{\i}sica; also at Instituto de Plasmas e Fus\~ao
Nuclear}\footnotetext{$^2$ Departamento de Engenharia Mec\^anica;
also at Centro de Ci\^encias e Tecnologias Aeron\'auticas e
Espaciais}\footnotetext{$^3$ Electronic address: {\tt
orfeu@cosmos.ist.utl.pt}} }

\author{F.~Francisco$^{2,4}$\footnotetext{$^4$ Electronic address: {\tt frederico.francisco@ist.utl.pt}}}

\author{P.~J.~S.~Gil$^{2,5}$ \footnotetext{$^5$ Electronic address: {\tt p.gil@dem.ist.utl.pt}}}

\author{J.~P\'aramos$^{1,6}$ \footnotetext{$^6$ Electronic address: {\tt jorge.paramos@ist.utl.pt}}}

\vskip 0.5cm

\affiliation{Instituto Superior T\'ecnico, \\ Avenida~Rovisco Pais 1,
1049-001 Lisboa, Portugal}

\vskip 0.2cm

\date{\today}

\begin{abstract}

We present a methodology based on point-like Lambertian sources that
enables one to perform a reliable and comprehensive estimate of the overall
thermally induced acceleration of the Pioneer 10 and 11 spacecraft.
We show, by developing a sensitivity analysis of the several
parameters of the model, that one may achieve a valuable insight on the
possible thermal origin of the so-called Pioneer anomaly.

\vskip 0.5cm

\end{abstract}

\pacs{07.87.+v, 24.10.Pa, 44.40.+a \hspace{2cm}Preprint
DF/IST-6.2008}

\maketitle

%%%%%%%%%%%%%%%%%%%%%%%%%%%%%%%%%% %

\section{Introduction}

\subsection{General Background}

The existence of an anomalous acceleration on the Pioneer 10 and 11
spacecrafts, sun-bound and with a magnitude of $a_\mr{Pio} \simeq (8.5
\pm 1.3) \times 10^{-10} ~\mr{m/s^2}$ has been put forward a decade
ago, using two independent code analyses \cite{JPL,Mark}. Attempts
to account for these phenomena as a result of a misestimation of the
systematic effects of thermal nature were first considered in Ref.\
\cite{Katz,Lou}. Possible additional contributions, ranging from electric
or magnetic forces, to mechanical effects or errors in the Doppler
tracking algorithms used, have all be shown to be unsuccessful.

Although initially dismissed, a much touted hypothesis for a physical explanation of the effect
lies in the reaction force due to thermal radiation arising from the
main bus compartment and the radiothermal generators (RTGs), either
directly pointing away from the sun, or reflected by the main
antenna dish. Clearly, an acceleration arising from the thermal
dissipation should present a similar secular trend as the RTGs
available power decay; regarding this point, one must note that
another analysis has shown that such a signature in the anomaly may
be found ({\it i.e.}, it also possesses statistical significance),
characterized by a linear decay with a time constant larger than 50
years \cite{Mark}: given the $\sim88$ years half-life of the
plutonium source in the radio-thermal generators, which should be
somewhat lowered due to degradation of the thermal coupling, this
still leaves room for thermal radiation to account for the Pioneer
anomaly. The latter is being thoroughly examined by groups within
the Pioneer collaboration team \cite{team,Toth}.

In what concerns other effects, one can safely disregard
electromagnetic forces, solar radiation and solar wind pressure as
the cause for the anomalous acceleration \cite{JPL}. Other sources
for anomalous effects have been discarded, including the possibility
that the Kuiper Belt's gravitational pull may give rise to the
reported acceleration; this would require an abnormally high mass
for this extended object, about two order of magnitude higher than
the commonly accepted value of $M_{\mr{Kuiper}} = 0.3
M_{\mr{Earth}}$ \cite{JPL,Vieira, Nietokuiper} (for a variety of
mass distribution models \cite{Vieira}).

The two Pioneer probes are following approximately opposite
hyperbolic trajectories away from the Solar System. The fact that
the same anomaly was found indicates a common origin to both
spacecraft. This prompts for an intriguing question: what is the
fundamental, and possibly new, physics behind this anomaly?

Many proposals have been advanced to explain the anomaly as a
previously undiscovered effect of new physics (see Ref.\
\cite{Paramos} and references therein, and also Refs.\
\cite{Reynaud,Moffat,Lobo}). However, before one seriously considers the
possibility for new physics, an unambiguous description of the
anomaly should be given. Unfortunately, the distances at which the
originally available Doppler measurements were conducted do not
allow for a clear discrimination of the direction of the
acceleration: in particular, it is still not possible to discern
between an acceleration towards the sun or the Earth, along the line
of sight. Ascertaining this would provide a relevant insight
concerning the origin of the anomaly: a line of action pointing
towards the sun would indicate a gravitational origin (since solar
radiation pressure is manifestly too low to account for the effect),
while a Earth-bound anomaly would hint at either a modified Doppler
effect (due to new physics affecting light propagation and causing
an effective blue shift) or an incorrect modeling of Doppler data,
possibly due to mismodeled Earth orientation parameters, incorrect
ephemerides estimates, Deep Space Network (DSN) and software clock drifts,
{\it i.e.}, an unaccounted systematic effect. An intriguing possibility
could be a ``congenital'' relationship between the Pioneer anomaly
and the so-called flyby anomaly \cite{flyby}. The anomalous acceleration may also lie along the spin axis of the spacecraft: this would indicate
that onboard, underestimated systematic effects to be held
responsible for it; finally, an anomaly along the velocity vector
would hint at some sort of drag effect.

Regarding the latter, it is worth stating that this additional drag
does not seem to be due to dark matter or dust distribution, since
these are currently known to a good accuracy, and yield much lower
effects. Conversely, one may ask what density should the environment
have, so that a $v^2$ dependent drag force would account for the
anomaly: a straightforward calculation shows that this should be of
order $10^{-19}~\mr{g/cm^3}$ (see, e.g., Ref.\ \cite{Vieira}); for
comparison, the density of interplanetary dust, arising from
hot-wind plasma \cite{Kimura}, is below $10^{-24}~\mr{g/cm^3}$; the
density of interstellar dust (directly measured by the Ulysses
spacecraft) is even smaller, at about $ 3 \times
10^{-26}~\mr{g/cm^3}$. Also, a modification of geodetical motion,
hinting at an extension of General Relativity, could also account
for a velocity dependent anomalous acceleration (see, e.g., Ref.\
\cite{zacuto} for a detailed discussion).

Furthermore, it is clear that a careful study of secular and spatial
trends should be carried out, aiming to relate with possible thermal
or engineering causes for the anomalous acceleration. The previously
available data is likely to refer to an insufficiently long mission
timespan, which does not allow for a clear discrimination of a
hypothetical variation of the anomaly; to overcome this difficulty,
recently recovered data of the full mission is being analyzed by
distinct groups within the Pioneer collaboration team, with several
approaches aiming to obtain convergent answers to the above
questions (see {\it e.g.} Ref. \cite{Agnes}).

Although initially disregarded, the issue of the Pioneer anomaly has
grown in and number of peer-reviewed publications, reflecting the
increasing concern of the physics community. The characterization of
any additional anomalous acceleration was part of the scientific
objectives of several mission proposals put forward to the recent
ESA Cosmic Vision 2015-2025 programme \cite{zacuto,missions};
unfortunately, these efforts were ill-fated, leaving the community
without the means to get a direct answer to this intriguing enigma.

\subsection{Previous Work}

\begin{table*}[ht] \begin{center} \caption{Error budget for the
Pioneer 10 and 11, taken from Ref. \cite{JPL}.}

\label{tableJPL}

%\vskip 20pt

\begin{tabular}{rlll}
\hline\hline
  Item & Description of error budget constituents  & Bias~~~~~             & Uncertainty \\
       &                                           & $10^{-8} ~\rm cm/s^2$ & $10^{-8}~\rm cm/s^2$ \\\hline
       &                                           &                       & \\
     1 & {\sf Systematics generated external to the spacecraft:} &         & \\
       & a) Solar radiation pressure and mass      & $+0.03$               & $\pm 0.01$\\
       & b) Solar wind                             &                       & $ \pm < 10^{-5}$ \\
       & c) Solar corona                           &                       & $ \pm 0.02$ \\
       & d) Electro-magnetic Lorentz forces        &                       & $\pm < 10^{-4}$ \\
       & e) Influence of the Kuiper belt's gravity &                       & $\pm 0.03$\\
       & f) Influence of the Earth orientation     &                       & $\pm 0.001$ \\
       & g) Mechanical and phase stability of DSN antennae       &         & $\pm < 0.001$\\
       & h) Phase stability and clocks             &                       & $\pm <0.001$ \\
       & i) DSN station location                   &                       & $\pm < 10^{-5}$ \\
       & j) Troposphere and ionosphere             &                       & $\pm < 0.001$ \\[10pt]
     2 & {\sf On-board generated systematics:}     &                       & \\
       & a) Radio beam reaction force              & $+1.10$               & $\pm 0.11$ \\
       & b) RTG heat reflected off the craft       & $-0.55$               & $\pm 0.55$ \\
       & c) Differential emissivity of the RTGs    &                       & $\pm 0.85$ \\
       & d) Non-isotropic radiative cooling of the spacecraft    &         & $\pm 0.48$\\
       & e) Expelled Helium produced within the RTGs             & $+0.15$ & $\pm 0.16$ \\
       & f) Gas leakage                            &                       & $\pm 0.56$ \\
       & g) Variation between spacecraft determinations          & $+0.17$ & $\pm 0.17$ \\[10pt]
     3 & {\sf Computational systematics:}          &                       & \\
       & a) Numerical stability of least-squares estimation      &         & $\pm0.02$\\
       & b) Accuracy of consistency/model tests    &                       & $\pm0.13$ \\
       & c) Mismodeling of maneuvers               &                       & $\pm 0.01$ \\
       & d) Mismodeling of the solar corona        &                       & $\pm 0.02$ \\
       & e) Annual/diurnal terms                   &                       & $\pm 0.32$\\[10pt]
\hline &                                           &                       & \\
       & Estimate of total bias/error              & $+0.90$               & $\pm 1.33$ \\
       &                                           &                       & \\
\hline\hline
\end{tabular} \end{center} \end{table*}

A clear assessment of several systematic contributions to the
overall acceleration may be found on Table~\ref{tableJPL}, extracted
from Ref.\ \cite{JPL}. These baseline figures give a good measure of
the different orders of magnitude of the various effects involved,
and show that they do not account for the reported anomaly. As it
turns out, unaccounted thermal effects are the most conspicuous
sources of a systematic effect. In Refs.\ \cite{Katz,Lou}, estimates were
performed for the heat dissipation of several spacecraft components,
and claimed that a combination of several sources could account for
the anomalous acceleration. In order to ascertain or disprove these
and other claims, a more recent and thorough study has carried out
the convoluted task of carefully modeling the Pioneer probes, in
order to reproduce all relevant thermal effects with a sufficient
accuracy \cite{Toth}; a similar, independent effort is being
undertaken by other groups within the Pioneer collaboration team.

Although still preliminary, these attempts seem to indicate that
thermal effects may account for up to one third of the total
magnitude of the reported anomaly \cite{Slava}. As we shall see this result is consistent with our own estimates which indicate that thermal effects can account from about 35\% to 67\% of the anomalous acceleration.  However, it is the
authors' opinion that the many parameter estimation and modeling
strategies available up to now somehow cloud the overall picture,
with the physical significance being hindered by the technical depth
of the thermal behavior reconstitution. For this reason, the present
work attempts to drift somewhat away from the full modeling of every
engineering detail, and directs its attention to the physical basis
of the aforementioned thermal behavior. This stated, it is clear
that our approach is a complementary tool to the current endeavors:
indeed, while a poorer modeling of specific details will reduce the
overall confidence of the obtained results, the added simplicity,
computational clarity and speed allow for a convenient and much
needed sensitivity analysis of the several relevant parameters.

In this paper, we present the main features and the first results of a
method based on point-like Lambertian sources. As we shall see, the
presented method is already compatible with previous studies;
further developments shall focus on a more detailed analysis of the
reflectivity effects, while still aiming at a good balance of model
simplicity, computational speed and physical realism.

\section{Source Distribution Method}

\subsection{Motivation and Rationale}

As discussed in the previous section, no definitive statements about
the origin of the anomaly can be put forward until its full
characterization. This justifies an intensive effort to recover and
analyze the full flight data, and to develop approaches to
understand the overall thermal behavior of the Pioneer probes, so to
measure any previously unaccounted thermal radiation effects and to
isolate, rule out, or constrain possibly remaining, yet unknown,
effects.

However, the authors feel that this pursuit should be countered with
an approach focusing on the physical effects directly relevant to
the understanding of the problem. The central issue is how thermal radiation
is emitted, and reabsorbed and/or reflected, by the external surfaces of
the spacecraft and what is the resultant reaction force. Hence,
instead of a complex finite elements model, that requires modeling
of the whole spacecraft, we propose to develop a faster, more
versatile approach based on a distribution of a few point-like
thermal sources, simulating the thermal radiation emitted from the
spacecraft, and analyzing the effect of radiation when emitted
directly to space or when reflected or absorbed by another surface of
the spacecraft. This approach is complementary to the ones based on finite element analyses and does not focus on
the inner behavior of each component or surface, but instead attempt
to isolate different contributions from the major constituents of
the vehicles, namely the RTGs, antenna dish, and main bus
compartment.

There are several arguments justifying the interest and the
effectiveness of the present approach. It is impossible to model the
Pioneer spacecraft in a very precise way: it was built decades ago,
accuracy of the blueprints or existing models is limited and the
precise properties and degradation or damage of the materials, after
decades in space, is unknown. This implies that even in the case of
a full model of the spacecraft educated guesses will have to be
done, limiting the accuracy of the obtained results.

The impossibility of reliably describing several key parameters
should also limit the accuracy of any conclusions derived from a
more encompassing approach. Specifically, the limited temperature
data (provided only by six sensors on the main bus and two sensors
on the RTGs) and poor knowledge of the optical properties of the
materials introduce substantial uncertainties in the final result,
whatever is the adopted strategy. Thus, it is clear that the total
and electrical power, which are well known, must be the fundamental
parameters for any analysis. As will be shown, our approach is based
on this principle.

Moreover, thermal radiation possibly contributing to the anomalous
acceleration depends on the external surfaces of the spacecraft and
how the total power (and temperature profile) is distributed among
them. The insulation of the spacecraft walls should limit the
gradient of the temperature along each of the main external surfaces
(except in special places as, e.g., the louvers, that can be modeled
as separate sources if required) and make all the modeling of the
details of the compartment unessential to address the problem. We
argue that small details and small gradients in temperature of the
spacecraft external surfaces will not affect the results
considerably since, as we will see, the results are not too much
affected by the number of point-like sources representing an
extended surface (keeping the power constant). A small number of
point-like sources can then be used to simulate any foreseen
temperature gradient along each surface or a small localized extra
source of radiation. From the thermal radiation point of view this
is similar to a unevenly distribution of power by the few point-like
sources representing the surface. Sensitivity analysis regarding the
details of the spacecraft: shape modeling, temperature gradients,
and total power emitted by each surface, can be then performed by
varying the power assigned to each individual source in a prescribed
way.

The Pioneer spacecraft is spin stabilized and any reaction force
component due to radiation will cancel away over time except in the
direction of the axis of rotation. Most of the small contributions
possibly not taken into account should be irrelevant since, due to
the geometry of the spacecraft, most of them are expected to be
normal to the axis of rotation. This effect can be verified through
the sensitivity analysis if slightly different radiation
distributions by the sources lead to similar values of the anomalous
acceleration, as expected. It should be noted that, as we are
modeling relatively large radiating surfaces as point-like sources,
the present model cannot provide too reliable information about the
total reaction torque induced by the thermal radiation into the
spacecraft.

It is clear that any study of this scope necessarily involves a
large number of assumptions and hypotheses. Therefore, it is
important to have the ability to quickly test a wide variety of
scenarios and reach unambiguous conclusions about their
plausibility: this sensitivity analysis is crucially facilitated by
the short computation time of the present method. In addition, the
simplicity of the formulation keeps the involved physics visible
throughout the entire process, allowing for scrutiny of every step.
Finally, we emphasize that the key goal of our effort is to perform
a wide spectrum study of the parameter space for several physical
properties relevant to the thermal modeling of the Pioneer probes.
Our approach, while less comprehensive than a finite element model,
allows for a direct interpretation of results, easy adaptability, as
well as rather short computation times.

Obviously, this endeavor would be incomplete if its self-consistency
could not be assessed. Thus, before tackling the more interesting,
physical case of the Pioneer anomaly, a set of test cases is
performed to ascertain the effectiveness of the method.
Furthermore, the choice for a point-like source approach should also
be verified; this may be achieved by increasing the number of
sources and observing the convergence of the relevant quantities and
results. If deemed satisfactory, one may safely assume
that continuous surfaces and components can be suitably modeled by
point-like sources, so to still reproduce the physical interplay
between them, and hence allow for an extrapolation to the Pioneer
vehicles.

\subsection{Physical Formulation}

\label{formulation}

Our method is based on a distribution of isotropic and Lambertian
point-like sources. If $W$ is the emitted power, the time-averaged
Poynting vector-field for an isotropic source located at
$(x_0,y_0,z_0)$ is given by

\beq \mathbf{S}_{\mr{iso}} = {W \over 4\pi}{(x - x_0, y - y_0, z -
z_0) \over \left[(x - x_0)^2 + (y - y_0)^2 + (z - z_0)^2
\right]^{3/2}} ~~. \eeq

\noindent In the case of a Lambertian source the intensity of the
radiation is proportional to the cosine of the angle with the normal

\beq \mathbf{S}_{\mr{Lamb}} = {W \cos \th \over \pi}{(x - x_0, y -
y_0, z - z_0) \over \left[(x - x_0)^2 + (y - y_0)^2 + (z - z_0)^2
\right]^{3/2}} ~~. \eeq

\noindent Typically, one uses isotropic sources to model point-like
emitters and Lambertian sources to model surfaces. The Poynting
vector field of the source distribution is, then, integrated over
the surfaces in order to obtain the amount of energy illuminating
these, and the force produced. The former is given by the
time-averaged Poynting vector flux

\beqa \label{eq3} E_{\mr{ilum}}& =&\int \mathbf{S} \cdot \mathbf{n} ~ d A = \\
\nonumber && \int \mathbf{S}(\mathbf{G}(s,t)) \cdot \left({\partial
\mathbf{G} \over \partial s} \times{\partial \mathbf{G} \over
\partial t} \right)~ d s\,d t  ~~, \eeqa

\noindent where the function ${\mathbf G(s,t)}$ parameterizes the
relevant surface. The radiation illuminating the surface will
produce a perpendicular force; integrating
this force, {\it i.e.}, the radiation pressure multiplied by the unitary
normal vector, will give us the total force acting upon the corresponding
surface. The radiation pressure is thus given by

\beq \label{prad} p_{\mr{rad}}={\al \over c} \mathbf{S} \cdot \mathbf{n}~~, \eeq

\noindent taking into account a radiation pressure coefficient $1
\leq \alpha \leq 2$. The case $\al=1$ corresponds to full absorption
while $\al = 2$ indicates full diffusive reflection.

There will also be a force acting on the source of the radiation;
this can be obtained by integrating the radiation pressure
multiplied by a normalized radial vector field along a generic
surface

\beq \label{emit}\mathbf{F}_{\mr{emit}}=\int {\mathbf{S} \cdot
\mathbf{n} \over c} {\mathbf{S} \over ||\mathbf{S}||} ~ dA ~~. \eeq

If the object in study has a reasonably complex geometry (such as
the Pioneer spacecraft) there will be shadows cast by the surfaces
that absorb and reflect the radiation. The shadowing effect of the
illuminated surfaces is calculated with this same expression and
then subtracted to the force obtained for the emitting surface.
Alternatively, one may use an integration surface that encompasses
the illuminated surfaces. The total result is the sum of all effects
$\mathbf{F}_i$ --- force on the emitting surface, shadows and
radiation pressure on the illuminated surfaces $
\mathbf{a}_{\mr{Th}}=\sum_i \mathbf{F}_i / m_{\mr{Pio}}$.

\subsection{Test Cases}

\label{test_cases}

In order to demonstrate the efficiency of the proposed method, we
define a set of test cases to assess the quality of our
approximation. The key point is the ability to adequately represent the thermal radiation
emitted from an extended surface by a small number of
point-like sources, as opposed to having many small
thermal radiating elements.

In the performed test cases, a square emitting surface with
$1~\mr{m^2}$ is considered. The three components of the force are
then computed: force on the emitting surface, shadow caused by
another surface at a given position, and radiation pressure on the
surface. We compare the results for different numbers of sources,
while maintaining the total power fixed. It is expected that the
result converges to the exact solution as the number of radiation
sources increases. Our study shows that one is able to get a
reliable error estimate even when using a small number of sources to
model a surface.

For a surface emitting radiation that does not illuminate other
surfaces, one finds that the force is perpendicular to the former and only
depends on the total emitted power. Using Eq.\ (\ref{emit}) with
Lambertian sources on a surface on the $0xy$ plane, one obtains a
force in the $z$-axis direction, of magnitude $(2/3)
W_{\mr{surf}}/c$.

\begin{figure}

\epsfxsize=8.5cm \epsffile{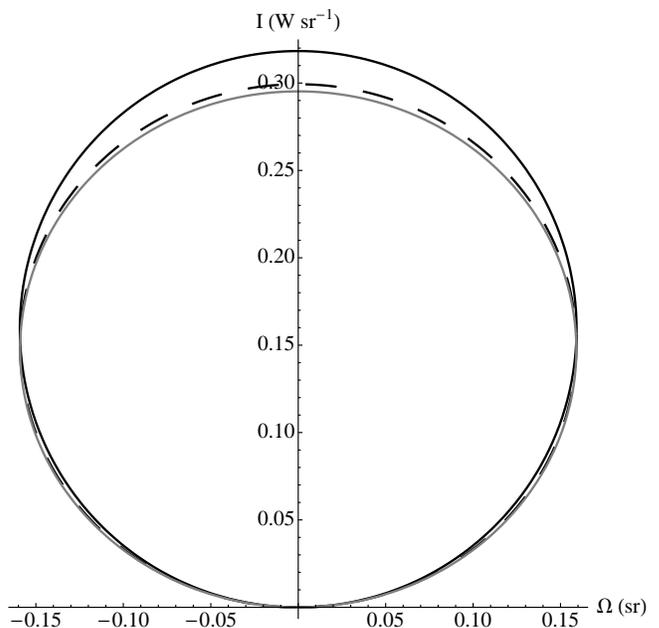} \caption{Polar
plot of the intensity variation with elevation of the radiation
emitted by a surface on the $0xy$ plane (solid angle $\Om$), when considering 1, 4 and 16 Lambertian sources (full, dashed and grey curves, respectively), maintaining the total power emitted by the surface constant at 1 $W$ (the curves for 64 or 144 sources overlap the one for 16 sources). The intensity at higher elevations (close to vertical) diminishes with the number of
sources, compensating the slight increase at the lower angles.}

\label{elevation}

\end{figure}

\begin{figure}

\epsfxsize=8.5cm \epsffile{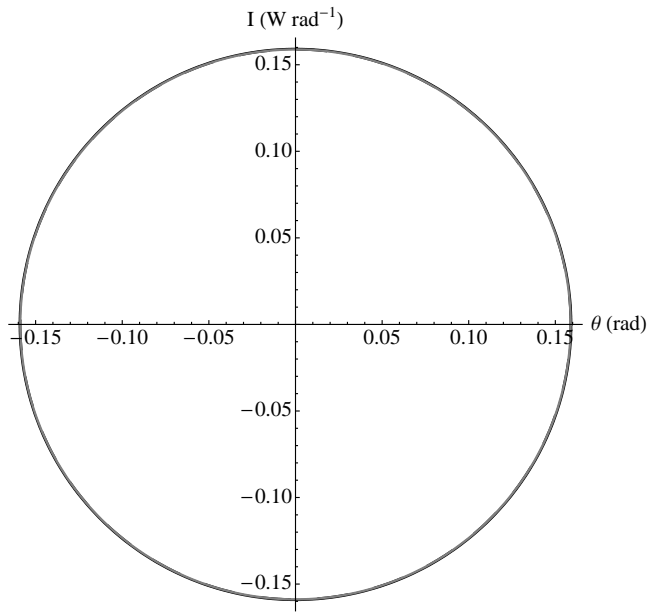} \caption{Same as
Fig.~\ref{elevation} but for intensity variation with azimuthal angle $\theta$. All lines are superimposed, confirming that the total power is kept constant.}

\label{azimuth}

\end{figure}

Computation of the shadow and pressure radiation on other surfaces
yields results that are not independent from the source
distribution. In order to acquire some sensibility on that
dependence, we plot the variation of the radiation intensity with
the elevation and the azimuth for 1, 4, 16, 64 and 144 source
meshes, as depicted in Figs. \ref{elevation} and \ref{azimuth}.

%%%%%%%%%%%%%%%%%%%%%%%%%%%%%%\~{O}%%%%%%%

%%%%%%%%%%%%%%%%%%%%%%%%%%%%%%%%%%%%%

%%%%%%%%%%%%%%%%%%%%%%%%%%%%%%\~{O}%%%%%%%

%%%%%%%%%%%%%%%%%%%%%%%%%%%%%%%%%%%%%

A visual inspection of the results indicate that, even for one
source, the maximum deviation occurs at the higher angles of elevation and is
less than 10\%. For the relevant angles for the Pioneer spacecraft
configuration, deviations will be considerably smaller. In order to
confirm this estimate, the force acting on a second $1~\mr{m^2}$
surface for several different positions is computed. A total of nine
representative configurations were considered, with different positions and tilt
angles, as summarized in Table \ref{testcase_configs}. The deviation
between the 1, 4, 16, 64 and 144 source meshes is then verified.

%%%%%%%%%%%%%%%%%%%%%%%%%%%%%%%%%%%%%%%%%

\begin{table}[ht] \begin{center}
\caption{Positions considered for the second surface in test cases.
The first (emitting) surface is in the $x-y$ plane centered at the
origin. Considered distances between both surfaces are typical for
the Pioneer spacecraft.}

\label{testcase_configs}

%\vskip 20pt

\begin{tabular}{cccc}

  \hline\hline
  Test case ~&~ Surface center position ~&~ Surface tilt angle  \\
  \# & ($\mr{m}$) & ($^\circ$) \\
  \hline
  $1$ & ($2$, $0$, $0.5$) & $90$  \\
  $2$ & ($2$, $0$, $1.5$) & $0$   \\
  $3$ & ($2$, $0$, $1.5$) & $30$  \\
  $4$ & ($2$, $0$, $1.5$) & $60$  \\
  $5$ & ($2$, $0$, $1.5$) & $90$  \\
  $6$ & ($1$, $0$, $2$)   & $0$   \\
  $7$ & ($1$, $0$, $2$)   & $30$  \\
  $8$ & ($1$, $0$, $2$)   & $60$  \\
  $9$ & ($1$, $0$, $2$)   & $90$  \\
  \hline\hline

\end{tabular} \end{center} \end{table}

%%%%%%%%%%%%%%%%%%%%%%%%%%%%%%%%%%%%%%%%%

\begin{figure}

\epsfxsize=4.5cm \epsffile{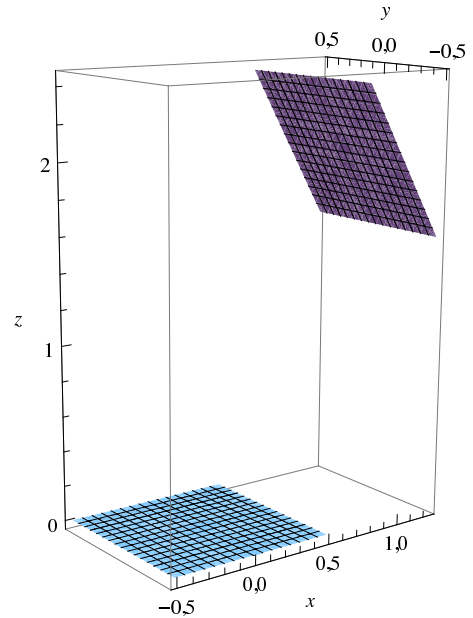} \caption{Geometry of
Test Case~8 (cf.\ Table~\ref{testcase_configs}): thermal emission
from a surface is simulated by a different number of Lambertian
sources evenly distributed on the surface, maintaining the total
power emitted constant, and the effect on a second surface is
observed. This is the test case where the highest variation with the
number os sources considered were obtained.}

\label{testfig8}

\end{figure}

Our study shows that the highest deviation occurs for Test Case~8, which confirms
our expectation, since the second surface is set at high elevation
from the emitting surface, as depicted in Fig.\ \ref{testfig8}. The
results in Table~\ref{table8} show a difference of
approximately 6\% between the force obtained with one source and the
results for the finer meshes (16, 64 and 144 sources). Nevertheless,
the latter are all within 0.5\% of each other, and the intermediate
4 source mesh has a deviation of only 1.5\%.

%%%%%%%%%%%%%%%%%%%%%%%%%%%%%%\~{O}%%%%%%%

%%%%%%%%%%%%%%%%%%%%%%%%%%%%%%%%%%%%%

%%%%%%%%%%%%%%%%%%%%%%%%%%%%%%%%%%%%%%%%%

\begin{table}[ht] \begin{center} \caption{Results for Test Case~8 (cf.\ Table~\ref{testcase_configs})
considering a total emission of $1~\mr{kW}$. As the number of
sources to represent the thermal emission of a surface change, the
resultant force components appearing by shadow on the secondary
surface remain almost the same.}

\label{table8}

%\vskip 20pt

\begin{tabular}{cccc}

  \hline\hline
  Sources ~&~ Energy flux ~&~ Force components $(x,y,z)$ \\
  \# & ($\mr{W}$) & ($10^{-7}~\mr{N}$) \\
  \hline
  $1$   & $45.53$ & ($2.016 $, 0, $2.083 $)  \\
  $4$   & $45.53$ & ($1.918 $, 0, $2.003 $)  \\
  $16$  & $45.53$ & ($1.895 $, 0, $1.984 $)  \\
  $64$  & $45.53$ & ($1.890 $, 0, $1.979 $)  \\
  $144$ & $45.53$ & ($1.889 $, 0, $1.978 $)  \\
  \hline\hline

\end{tabular} \end{center} \end{table}

%%%%%%%%%%%%%%%%%%%%%%%%%%%%%%%%%%%%%%%%%

\begin{figure}

\epsfxsize=8.5cm \epsffile{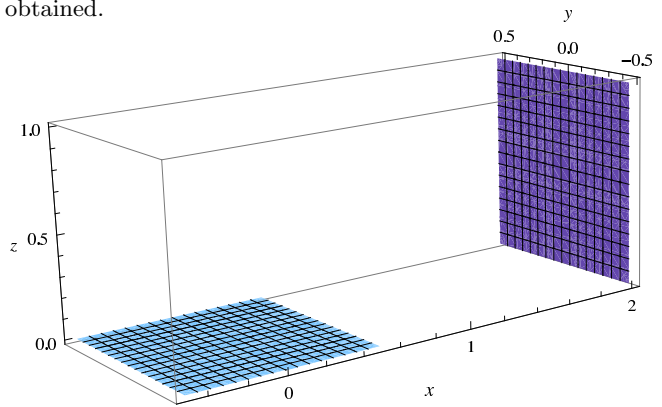} \caption{Same as
Fig.~\ref{testfig8}, for Test Case~1.}

\label{testfig1}

\end{figure}

For the typical angles of the Pioneer probe's configuration, one may
take as figure of merit Test Cases~1 and 3. For the first case,
depicted in Fig.~\ref{testfig1}, the radiation pressure and shadow
yield the results shown in Table~\ref{table1}. The analysis of these
results shows that, for 16, 64 and 144 sources, the variation in the
energy flux and force is less than 0.5\%. In addition to that, the difference when compared with the results from finer meshes is less than 5\% for 1 source
and less than 1.5\% for a 4 source mesh. The results in Table
\ref{table2} show, for Test Case 3, a variation of less than 5\%
between the results for 1 source and 144 sources. The convergence
is, as in both previous cases, achieved for the 16, 64 and 144
source meshes, with a variation of less than 0.25\%.

%%%%%%%%%%%%%%%%%%%%%%%%%%%%%%\~{O}%%%%%%%

%%%%%%%%%%%%%%%%%%%%%%%%%%%%%%%%%%%%%

%%%%%%%%%%%%%%%%%%%%%%%%%%%%%%%%%%%%%%%%%

\begin{table}[ht] \begin{center} \caption{Same as Table~\ref{table8}, for Test Case~1.}

\label{table1}

%\vskip 20pt

\begin{tabular}{cccc}

  \hline\hline
  Sources ~&~ Energy flux ~&~ Force components $(x,y,z)$ \\
  \# & ($\mr{W}$) & ($10^{-7}~\mr{N}$) \\
  \hline
  $1$   & $15.34$ & ($0.9300 $, 0, $0.1514 $)  \\
  $4$   & $15.92$ & ($1.028 $, 0, $0.1638 $)  \\
  $16$  & $16.09$ & ($1.038 $, 0, $0.1675 $)  \\
  $64$  & $16.13$ & ($1.040 $, 0, $0.1684 $)  \\
  $144$ & $16.14$ & ($1.041 $, 0, $0.1686 $)  \\
  \hline\hline

\end{tabular} \end{center} \end{table}

%%%%%%%%%%%%%%%%%%%%%%%%%%%%%%%%%%%%%%%%%

%%%%%%%%%%%%%%%%%%%%%%%%%%%%%%%%%%%%%%%%%

\begin{table}[ht] \begin{center} \caption{Same as Table~\ref{table8}, for Test Case~2.}

\label{table2}

%\vskip 20pt

\begin{tabular}{cccc}

  \hline\hline
  Sources ~&~ Energy flux ~&~ Force components $(x,y,z)$ \\
  \#    & ($\mr{W}$)   & ($10^{-7}~\mr{N}$) \\
  \hline
  $1$   & $19.20$ & ($0.4952 $, 0, $1.037 $)  \\
  $4$   & $19.83$ & ($0.5032 $, 0, $1.082 $)  \\
  $16$  & $19.99$ & ($0.5050 $, 0, $1.093 $)  \\
  $64$  & $20.03$ & ($0.5054 $, 0, $1.096 $)  \\
  $144$ & $20.04$ & ($0.5055 $, 0, $1.096 $)  \\
  \hline\hline

\end{tabular} \end{center} \end{table}

%%%%%%%%%%%%%%%%%%%%%%%%%%%%%%%%%%%%%%%%%

For all test cases examined, the convergence of the results occurs
at a similar pace and yields, for all cases,  similar small deviations. Ultimately,
we conclude that a 4 source mesh, with deviations around 1.5\%,
would be adequate for the desired balance between precision and
simplicity. These results provide a fairly good illustration of the
power of our method and how well we can estimate the radiation
effects on the Pioneer probes. In particular, the deviation is
always well below 10\%, even with the roughest simplifications
allowed by the chosen method. One may  then conclude that, for the
scales and geometry involved in the Pioneer anomaly problem, the
source distribution method is, not only consistent and convergent,
but that it provides a very satisfactory estimate of the thermal
radiation effects, even considering all uncertainties involved.

Finally, after analysing the convergence of the method, we have also considered two additional test cases to assess the effect of ignoring some surface features, such as the equipment attached to the external walls of the spacecraft. These results indicate that, unless large temperature gradients are present, no significant errors will arise from considering flat surfaces and not taking into account all the details of the spacecraft.

\section{Thermal Radiation Model of the Pioneer Spacecraft}

\subsection{Geometry}

\label{geometry}

The problem of modeling the Pioneer spacecraft can be considerably
simplified with some sensible hypotheses. The first and most
important one rests upon the fact that the probes are spin
stabilized. Since it is also assumed that the probe is in a
steady-state thermal equilibrium through out most of their journey,
the time-averaged radial components of any force generated by
anisotropic radiation will be negligible. In addition, the probe's
axis of rotation (taken as the $z$-axis) is approximately pointing
towards Earth, which is also the approximate direction of the anomalous
acceleration.

\begin{figure}

\epsfxsize=6cm \epsffile{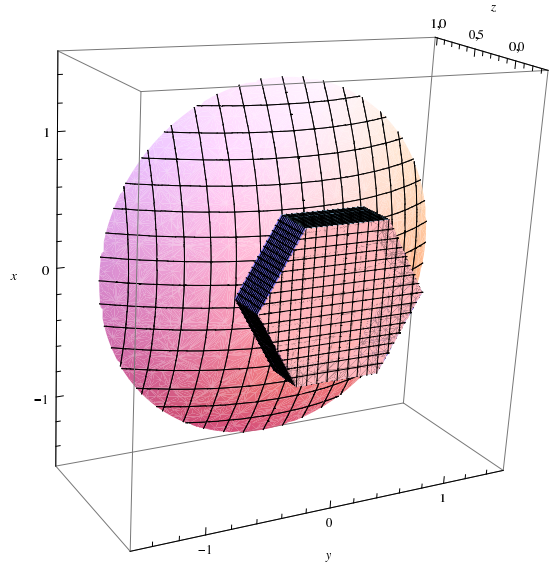} \caption{Pioneer
spacecraft model geometry considered in calculations, back view:
high gain antenna and hexagonal main bus compartment.}

\label{pioneer}

\end{figure}

\begin{figure*}

\epsfxsize=16cm \epsffile{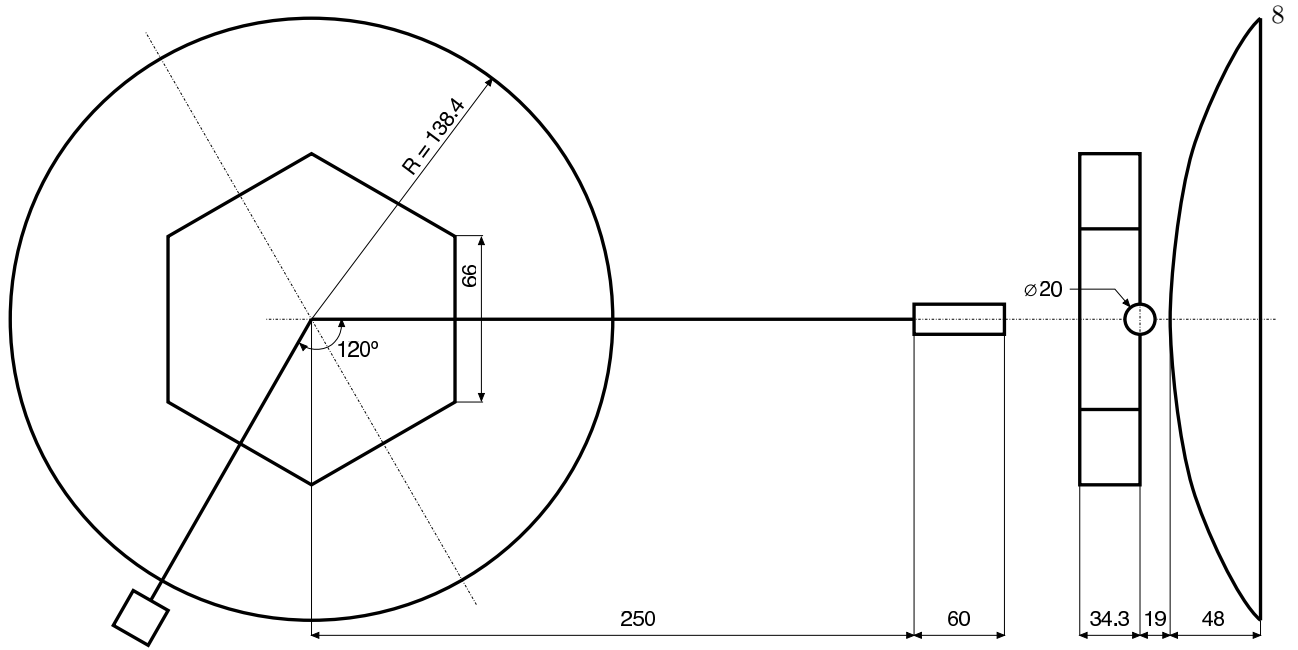} \caption{Schematics of our simplified model of the Pioneer
spacecraft, with relevant dimensions (in $\mr{cm}$); second RTG and truss are not represented to scale, for convenience. Lateral view indicates the relative position of the RTGs, box compartment and the gap between the latter and the high gain antenna.}

\label{esquema}

\end{figure*}

In this study, we consider a simplified version of the spacecraft
geometry, which retains only its most important features, as
depicted in Figs.~\ref{pioneer} and ~\ref{esquema}. Our model considers the RTGs, a prismatic equipment compartment and the antenna ---
a paraboloid, parametrized by the function ${\mathbf G}(s,t) = \left( s,~t,~a(s^2+t^2) \right) $, with a parabolic coefficient $a= 0.25~\mr{m}^{-1}$ (c.f. Eq. (\ref{eq3})).
Dimensions are taken from the available Pioneer technical drawings.
Our results are obtained through the integration of the emissions of
the RTG and lateral walls of the equipment compartment along the
visible portion of the antenna. Note that radiation emitted from the
front surface of the Pioneer cannot be reflected by other surfaces
and will be counted as a whole. The surface of the compartment
facing the antenna will be discarded for now as its contribution is fairly small for obvious geometric reasons: escaping radiation will
be attenuated by multiple reflections between these two components and will be mainly in the
radial direction, not contributing significantly to the anomalous
acceleration. The antenna itself is expected to have a very low temperature ($\sim 70~\mr{K}$) with an approximately uniform distribution, not only axially, but also considering the front and back surfaces of the paraboloid (as visible in the
results from Ref.\ \cite{Slava}); therefore, its contribution can be regarded as negligible, with the surface acting solely as a reflector for the incoming radiation.

%%%%%%%%%%%%%%%%%%%%%%%%%%%%%%%%%%%%%

%%%%%%%%%%%%%%%%%%%%%%%%%%%%%%%%%%%%%

%%%%%%%%%%%%%%%%%%%%%%%%%%%%%%%%%%%%%

%%%%%%%%%%%%%%%%%%%%%%%%%%%%%%%%%%%%%

As we shall see, this simplified model captures the most important
contributions to the thermal reaction force. The RTGs and the main
equipment compartment are actually responsible for most of the
emitted thermal radiation. In the case of the
equipment compartment, the most important contribution is from the
louvers located in the front wall (facing away from the sun) with
consequences for the total power distribution.

\subsection{Point-like Source Distribution}

\label{source_distrib}

In order to estimate the thermal effects, a separate analysis
of the three main contributions must be performed. The front wall of
the probe, where the louvers are located, is perpendicular to the
axis of rotation: its contribution corresponds to a force $(2/3)
W_{\mr{front}}/c$ pointing in the sun-ward direction along the
probe's axis. The contribution from the side walls of the main
compartment is obtained from the integration of the shadow and
radiation pressure components along the antenna. The shadow of the
RTGs was neglected since they are small, relatively distant, and
most of its effect would be in the radial direction. Following an
approach similar to the one used in the test cases, in order to
verify the convergence of the result, the integration is performed
for an increasing number of sources. The results converge fairly
quickly and the deviations are all below 2.5\%, confirming the
consistency previously demonstrated in our test cases. The obtained
values show that between 16.8\% to 17.3\% of the power emitted from
the side walls of the compartment is converted into a sun-ward
thrust along the $z$-axis.

It is also important to verify how the results are affected by a
non-uniform temperature distribution. This is simulated by varying
the relative power of the point-like sources in each surface,
keeping the total power attributed to the surface constant. A
variation of 20\% in power between sources (simulating a 5\%
temperature variation) gives no significant changes in the final
result --- with relative differences smaller than 1\%.

Finally, the RTG contribution is computed through two different
models. The first, simpler scenario, mimics each RTG with a single
isotropic source. In this case, the point-like source has the whole
power of the RTG. In the second model, the cylindric shape of the
RTG is taken into account and a Lambertian source is placed at each
base. Actually, it is only necessary to consider the source facing
towards the centre of the spacecraft, as the remaining RTG radiation
will be emitted radially and its time-averaged contribution
vanishes. In this case, the Lambertian source has a certain amount
of the total RTG power, as discussed in the following sections.
Depending on the model considered, either 1.9\% of the total power
or 12.7\% of the power emitted from the base of the cylinder
(equivalent to approx.\ 2\% of total RTG power, if the temperature is
uniform) is converted into thrust.

These preliminary results do not take into account diffusive
reflection, as allowed by Eq.\ \ref{prad}. In the subsequent
section, more accurate results will be presented and discussed.

\subsection{Available Power}

The available power on the Pioneer spacecraft is one of the few
measured or inferred parameters that is reasonably well known. In
addition, it is physically more consistent to consider the power instead of
the temperature readings as it is the independent variable from which all
estimates of the resulting thermal effects are derived. Of course,
the energy balance to the spacecraft in steady-state conditions
relates the temperature $T_i$ of a surface $i$ with the power budget
of the spacecraft

\beq \dot{E}_{\mr{absorb}}+\dot{E}_{\mr{gen}}=\sum_i A_i \epsilon_i
\sigma T_i^4~~, \eeq

\noindent where $A_i$ are the relevant areas and $\ep_i$ the
emissivity of each surface $i$.

Notice that, since the optical properties of the surfaces, as well as
their evolution with time, are not well known, temperature estimates
are quite uncertain. Furthermore, as a variation in the emissivity
implies the violation of the conservation of energy, a new solution
for the temperature must be obtained iteratively for each different
set of optical properties used so to maintain the correct power.

All the power generated onboard the probes comes from the two
plutonium-238 RTGs. Given that just a fraction of the generated heat
is converted into electricity, the remaining power is dissipated as
heat. There will be some conduction of this heat through the truss
to the central compartment, however, considering the small section
of this structure, it is reasonable to admit that it will have a
reduced impact on the total RTG radiated power\footnote{Assuming the RTGs trusses as hollow cylinders, with radius $1~\mr{cm}$ and a temperature gradient of $30~\mr{K/m}$, one obtains about $\lesssim 4~\mr{W}$ of dissipated power, which is clearly negligible.}. It is, thus, considered that all of the RTG thermal power is dissipated as
radiation from the RTG itself.

The electrical power is consumed by the various instruments located
in the main compartment, despite a considerable portion of it being
used in radio transmissions from the high gain antenna. As mentioned
in Ref.\ \cite{JPL}, the total RTG thermal power at launch was
2580~W, producing 160~W of electrical power. This means that, at
launch, approximately 2420~W of thermal power has been dissipated by
the RTGs. Taking into account the plutonium decay with a half-life
of 87.74 years, the total on-board power variation with time (in
years) is given by

\beq W_{\mr{tot}}(t)= 2580 \exp \left(-{\frac{t \ln 2
}{87.74}}\right)~~. \eeq

\noindent Telemetry data reveals that the electrical power decays at a faster
rate than the plutonium radioactive decay; in the latest stages of
the mission, about 65~W were available. Most of the electrical power
is dissipated inside the main compartment. The electrical heat in
the spacecraft body was around 120~W at launch, dropping to less
than 60~W at the latest stages of the mission \cite{Toth}, following
an approximate exponential decay with a half-life of about 24 years.
This difference in decay rates is mostly attributable to
thermocouple degradation.

\section{Results and Discussion}

\subsection{Order of Magnitude Analysis}

Before undertaking a more rigorous numerical estimate, one may use
the results described above to perform a preliminary order of
magnitude analysis. This allows one to obtain a concrete figure of
merit for the overall acceleration arising from thermal effects,
which can be compared with the $a_\mr{Pio} \sim 10^{-9}~\mr{m/s^2}$
scale of the Pioneer anomaly.

From the spacecraft specifications, one has a total mass
$m_{\mr{Pio}} \sim 230~\mr{kg}$, and separate RTG and equipment
compartment powers $W_{\mr{RTG}} \sim 2 ~\mr{kW}$ and
$W_{\mr{equip}} \sim 100 ~\mr{W}$. As already discussed, the
integration of the emissions of the RTG and instrument compartment
indicate the proportion of emitted power that is effectively
converted into thrust. If we consider the simpler model discussed is
section \ref{source_distrib} and the power emitted from each surface
proportional to its area (equivalent to assuming uniform temperature
and emissivity in the RTGs and equipment compartment), we obtain

\beqa F_{\mr{RTG}}& \sim &2 \times 10^{-2} {W_{\mr{RTG}} \over c}~~,\\
\nonumber F_{\mr{sides}} &\sim& 10^{-1} {W_{\mr{equip}} \over
c}~~, \\ \nonumber F_{\mr{front}} &\sim& 2 \times 10^{-1} {W_{\mr{equip}} \over
c}~~. \eeqa

One can easily estimate the acceleration of the spacecraft due to
the thermal effects arising from the power dissipation of the RTGs
and equipment compartment:

\beqa a_{\mr{RTG}} & \sim & 2 \times 10^{-2} {W_{\mr{RTG}} \over
m_{\mr{Pio}}c} \sim 6 \times 10^{-10}~\mr{m/s^2}~~,\\ \nonumber
a_{\mr{equip}} & \sim & 3 \times 10^{-1} {W_{\mr{equip}} \over
m_{\mr{Pio}}c} \sim 4.4 \times 10^{-10}~\mr{m/s^2}~~. \eeqa

\noindent This clearly indicates that both contributions are relevant to
account for the reported anomalous acceleration of the Pioneer
probes. Furthermore, it also shows that the RTGs and the instrument
compartment yield similar thermal effects, so that one cannot focus
solely on one of these sources when modeling the spacecraft (this
had already been revealed by the analysis of Ref.\ \cite{Toth}).

\subsection{Thermal Force Estimate}

Encouraged by the estimate outlined above, one may now proceed  with
a more thorough evaluation of the existing thermal effects, using
our point-like source modeling.

In this section we shall use a model with 4 sources in each side
panel of the equipment compartment and Lambertian sources at the
bases of the RTGs, as discussed in section \ref{source_distrib}. We
believe this model gives us the best compromise between accuracy and
computation time --- the deviation of the source distribution
relative to the finer meshes is less than 0.5 \%. Integrating the radiation pressure and shadow components using the methodology presented in section \ref{formulation} and extracting the axial component, we obtain an expression that yields the thermal acceleration

\beq  \label{acceleration} a_{\mr{Th}} = { \left( 0.168 W_{\mr{sides}} + {2 \over 3} W_{\mr{front}} +
0.128 W_{\mr{RTGb}} \right) \over m_{\mr{Pio}}c}~~, \eeq

\noindent where $W_{\mr{sides}}$ and $W_{\mr{front}}$ are the powers
emitted from the side panels and front of the equipment compartment
and $W_{\mr{RTGb}}$ is the power emitted from the base of the RTG
facing the centre of the spacecraft. Remaining contributions are much smaller, as discussed in sections \ref{geometry} and \ref{source_distrib}.

A critical analysis of this expression, bearing in mind the spacecraft geometry, reveals that all considered contributions yield a sun-ward acceleration: the $W_{\mr{front}}$ component radiates directly in a direction away from the sun, while the other two components $W_{\mr{sides}}$ and $W_{\mr{RTGb}}$ radiate laterally, illuminating the high gain antenna --- which will yield a significant shadow and radiation pressure. The question now resides in correctly estimating each one of these
powers. We shall consider the 1998 readings, as found in the graph of
Ref.\ \cite{Toth}, namely: $W_{\mr{RTG}}=2050 ~\mr{W}$ and
$W_{\mr{equip}}=58 ~\mr{W}$. These are the dissipated thermal powers
at the RTG and equipment compartment.

The simplest scenario, with uniform temperature and
optical properties (emissions proportional to the surface area, as
in the previous section), leads to

\beqa W_{\mr{sides}}&=& 21.75 ~ \mr{W}~~, \\ \nonumber W_{\mr{front}}&=&18.12~ \mr{W} ~~,\\
\nonumber W_{\mr{RTGb}}& =&41.11~ \mr{W}~~,\eeqa

\noindent yielding an acceleration $a_{\mr{Th}}=3.05 \times
10^{-10}~\mr{m/s^2}$. This
amounts to about 35\% of the anomalous acceleration.
However, it is wise to undertake a critical analysis of this figure:
considering the available temperature maps of Refs.\
\cite{Toth,Slava}, we see that the temperature anisotropies along
the sides of the equipment compartment fall within the tested cases,
as discussed in section \ref{source_distrib}. However, the RTG
temperature distribution should deserve further attention, as there
are significant temperature changes between the wall of the
cylinder, the bases and the fins. In addition, it is expected that
the front wall of the equipment compartment will have a larger
contribution than the side walls, due to the presence of the
louvers.

\begin{figure}

\epsfxsize=8.5cm \epsffile{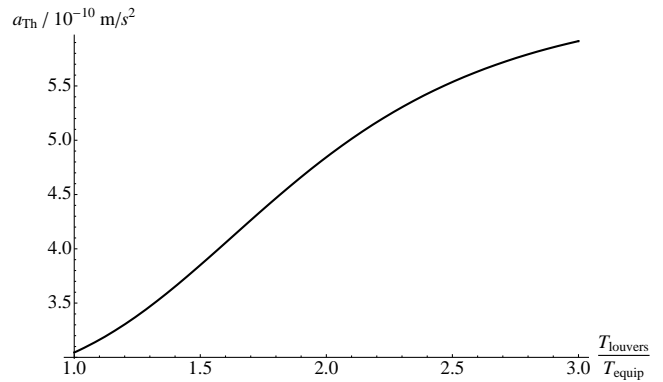}
\caption{Variation of the resulting acceleration with the
temperature ratio between the louvers and the equipment platform,
considering similar emissivities for both multi-layer insulations.}

\label{graphTlouvers}

\end{figure}

\begin{figure}

\epsfxsize=8.5cm \epsffile{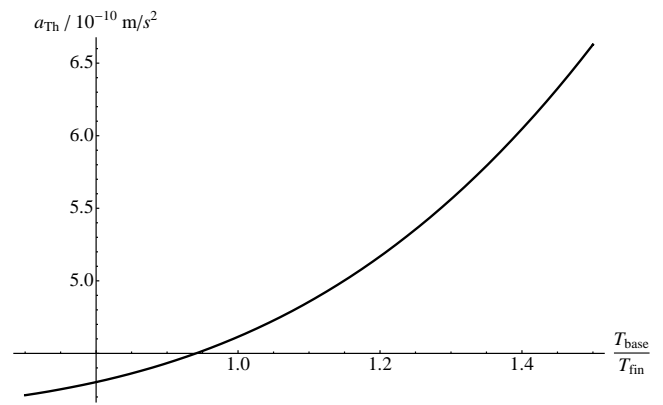} \caption{Variation
of the resulting acceleration with the temperature ratio between the
base of the RTG cylinder and the fin temperature.}

\label{graphTrtg}

\end{figure}

Taking these considerations into account, one can analyze the
variation of the emitted power in the louvers and at the base of the
RTG, since these are the two critical parameters in the
calculation. If we consider that the louvers are closed and have a
similar emissivity to the rest of the equipment platform, we can
plot the variation of the acceleration with the temperature ratio
between the louvers and the mean temperature of the platform, while
keeping the total power constant. This is depicted in Fig.\
\ref{graphTlouvers}. One can perform a similar analysis for the
RTGs, considering the ratio between the temperatures at the base of
the cylinder and the fins (Fig.\ \ref{graphTrtg}).

%%%%%%%%%%%%%%%%%%%%%%%%%%%%%%%%%%%%%

%%%%%%%%%%%%%%%%%%%%%%%%%%%%%%%%%%%%%

%%%%%%%%%%%%%%%%%%%%%%%%%%%%%%%%%%%%%

%%%%%%%%%%%%%%%%%%%%%%%%%%%%%%%%%%%%%

Figs.\ \ref{graphTlouvers} and \ref{graphTrtg} are illustrative of
the main strength of our method: it allows for a fairly quick and
accurate analysis of the dependence of the final result on different
parameters. Through Eq.\ (\ref{acceleration}) and sensible variation
of the power parameters, one can match temperature readings and
consider hypotheses for the variation of the optical properties.

We can now perform a second estimate considering the RTG cylinder
bases and wall as having a 15\% and 30\% higher temperature than the
fins, respectively. Assuming also that the closed louvers have
similar emissivities, although a 100\% higher temperature than the
rest of the equipment compartment could be possible, one obtains the
following values for the powers:

\beqa W_{\mr{sides}}&=& 9.97~\mr{W}~~,\\ \nonumber  W_{\mr{front}}&=&39.71~\mr{W} ~~,\\
\nonumber W_{\mr{RTGb}}&=&49.67~\mr{W}~~. \eeqa

\noindent In this case, one can account for 57\% of the anomalous
acceleration, that is, $a_{\mr{Th}}=5.00 \times
10^{-10}~\mr{m/s^2}$.

So far, our results do not consider reflections, {\it i.e.}, full
absorption of the radiation by the illuminated surfaces. In this
study, we shall introduce diffusive reflection by assigning a value
to the $\al$ parameter in Eq.\ (\ref{prad}). For the kind of
aluminum used in the construction of the antenna, the reflectivity
is, typically, around 80\% for the relevant wavelengths, yielding
$\al=1.8$. For the multi-layer insulation of the equipment platform,
a value of $\al=1.7$ is considered. In these conditions, the
illumination factors in Eq.\ (\ref{acceleration}) are modified to
account for the reflection. With the same temperature conditions as
in the previous case, the resulting acceleration is
$a_{\mr{Th}}=5.75 \times 10^{-10}~\mr{m/s^2}$ --- approximately two thirds of the anomalous acceleration.

The results presented in this section give us a fairly good idea of
the changes involved when considering different hypotheses and
parameters. The three discussed scenarios here illustrate how one
can use our method to identify the most sensitive parameters and
quickly assess the effect of the existing uncertainties, suggesting where models must be refined in order to increase confidence in results.

\section{Conclusions}

In this work we have developed a method to account for the
acceleration of the Pioneer spacecraft due to thermal effects,
based on point-like Lambertian sources. The flexibility and computation simplicity of our method allow for a reliable and fast estimate of the acceleration due to the various thermal contributions of the spacecraft components. This is sharply contrasting with the complexity and computationally demanding nature of the finite element analysis. Our methodology is potentially useful for a thorough parametric study of the various thermal contributions, as discussed in sections III and IV.

Our method allows for a reasonable degree of
accuracy and the numerical error estimates provided by the numerical
calculation package are of the order of $10^{-14}$ or less, while
the approximation of the geometry with point-like sources results in
a deviation of less than 1\%, as discussed in sections
\ref{test_cases} and \ref{source_distrib}. This should not be
understood as an indication of the accuracy of the resulting
accelerations, when compared to the reported case of the Pioneer
anomaly, but as a measure of self-consistency and reliability of the
developed method --- which should be expanded to model the physical
system of the Pioneer spacecrafts more closely, while maintaining
the desired flexibility and computational speed.

We find, after
identifying the main contributions for the power of the various
components of the spacecraft (RTGs, antenna and equipment bus
compartment), figures ranging from 35\% to 57\% of the
anomalous acceleration disregarding reflection. Inclusion of reflection implies that one can account for about 67\% of the anomaly. 

The natural continuation of this work will involve the refinement of the
geometrical modeling, including the specular component reflection.
In addition, and possibly more relevantly, we aim to pursue the identification of
parameters that most significantly affect the final result --- namely
temperatures, emissivities and reflectivities of the various
components, such as the louvers and the RTG case. In any case, our analysis does
achieve a reasonable level of agreement with other thermal models
based on finite element methods \cite{Toth,Slava}.

%%%%%%%%%%%%%%%%%%%%%%%%%%%%%%%%%%%%%%

\begin{acknowledgments}

\samepage \noindent This work is partially supported by the Programa
Dinamizador de Ci\^{e}ncia e Tecnologia do Espa\c{c}o of the FCT ---
Funda\c{c}\~{a}o para a Ci\^{e}ncia e Tecnologia (Portuguese Agency), under the
project PDCTE/FNU/50415/2003, and partially written while attending
the third Pioneer Anomaly Group Meeting at the International Space
Science Institute (ISSI) at Bern, from 19 to 23 of February 2008.
The authors would like to thank ISSI and its staff, for hosting the
group's meeting and accommodating for logistic requirements. The
work of JP is sponsored by the FCT under the grant BPD 23287/2005.

\end{acknowledgments}

%%%%%%%%%%%%%%%%%%%%%%%%%%%%%%%%%%%%%%

\end{document}